\documentclass[pra,a4paper,aps,twocolumn,nofootinbib,superscriptaddress,10pt,showkeys, longbibliography]{revtex4-1}
\usepackage[T1]{fontenc}

\usepackage[dvipsnames]{xcolor}
\usepackage{amsmath,amsthm,amsfonts,graphicx,times,xfrac,booktabs,mathtools,enumitem,xr,subfigure,amssymb,bbm,verbatim,appendix,placeins}
\usepackage[unicode=true,bookmarks=true,bookmarksnumbered=false,bookmarksopen=false,breaklinks=false,pdfborder={0 0 1}, backref=false,colorlinks=true]{hyperref}
\usepackage{orcidlink}
\usepackage{bm}
\usetikzlibrary {shadows.blur, decorations.pathreplacing, arrows, positioning} 

\usepackage{graphicx}
\usepackage{siunitx}
\usepackage{placeins}
\usepackage{makecell}

\renewcommand*{\url}[1]{\href{#1}{#1}}

\makeatletter
\theoremstyle{plain}

\theoremstyle{plain}

\theoremstyle{plain}

\theoremstyle{remark}
\newtheorem*{rem*}{\protect\remarkname}
\theoremstyle{plain}

\theoremstyle{plain}

\theoremstyle{definition}

\theoremstyle{plain}

\theoremstyle{plain}
\newtheorem*{thm*}{\protect\theoremname}
\theoremstyle{plain}
\newtheorem*{lem*}{\protect\lemmaname}
 
\providecommand{\propositionname}{Proposition}
\providecommand{\theoremname}{Theorem}
\providecommand{\lemmaname}{Lemma}
\providecommand{\remarkname}{Remark}
\providecommand{\conjecturename}{Conjecture}
\providecommand{\definitionname}{Definition}
\providecommand{\corollaryname}{Corollary}
\providecommand{\observationname}{Observation}
\allowdisplaybreaks

\def\bra#1{\langle{#1}\vert}
\def\ket#1{\vert{#1}\rangle}

\def\BraVert{e.g.,roup\,\mid\,\bgroup}

\newcommand{\tr}{\mathrm{tr}}

\newcommand{\SA}{{\Sys\!\Anc}}
\newcommand{\rhoSA}{\rho^\SA}
\newcommand{\rhoSAzero}{\rho^\SA_0}
\newcommand{\rhoSAone}{\rho^\SA_{t_1}}
\newcommand{\rhoSAtwo}{\rho^\SA_{t_2}}
\newcommand{\rhoSAt}{\rho^\SA_{t}}

\newcommand{\id}{\mathbbm{1}}

\let\oldaddcontentsline\addcontentsline

\newcommand{\starttocentries}{\let\addcontentsline\oldaddcontentsline}

\newcommand{\cpt}{\mathcal{E}}
\newcommand{\map}{\Phi}
\newcommand{\mapser}{\Phi_{\dt}}

\newcommand{\dt}{\delta t}
\newcommand{\conc}{\mathcal{C}}
\newcommand{\Sys}{\mathcal{S}} 
\newcommand{\Anc}{\mathcal{A}} 
\newcommand{\Env}{\textrm{E}}
\newcommand{\Dyn}{\mathcal{D}}

\newcommand{\g}{g}
\newcommand{\gdt}{g \delta t}
\newcommand{\ug}{U_{\delta}}

\def\e{\ensuremath{\mathrm{e}}}
\def\iu{\ensuremath{\mathrm{i}}}
\def\i{\iu}

\theoremstyle{definition}

\newcommand{\rzph}{\gate{R_Z(\frac{\pi}{2})}}
\newcommand{\rzmph}{\gate{R_Z(-\frac{\pi}{2})}}
\newcommand{\sx}{\gate{\sqrt{X}}}
\newcommand{\ecr}{\gate[2, disable auto height]{\verticaltext{ECR}}}
\newcommand{\rzpf}{\gate{R_Z(\frac{\pi}{4})}}
\newcommand{\rzmpf}{\gate{R_Z(-\frac{\pi}{4})}}
\newcommand{\rztpf}{\gate{R_Z(\frac{3\pi}{4})}}
\newcommand{\rzmtpf}{\gate{R_Z(-\frac{3\pi}{4})}}
\newcommand{\rzmp}{\gate{R_Z(\frac{\pi}{4})}}

\usepackage{tikz}
\usetikzlibrary {shadows.blur, decorations.pathreplacing, arrows, positioning} 
\usetikzlibrary{quantikz2}

\usepackage{afterpage}

\usepackage{tikz}
\usetikzlibrary{patterns}
\usetikzlibrary{patterns.meta}

\tikzdeclarepattern{
  name=hatch,
  parameters={\hatchsize,\hatchangle,\hatchlinewidth},
  bounding box={(-.1pt,-.1pt) and (\hatchsize+.1pt,\hatchsize+.1pt)},
  tile size={(\hatchsize,\hatchsize)},
  tile transformation={rotate=\hatchangle},
  defaults={
    hatch size/.store in=\hatchsize,hatch size=5pt,
    hatch angle/.store in=\hatchangle,hatch angle=0,
    hatch linewidth/.store in=\hatchlinewidth,hatch linewidth=.4pt,
  },
  code={
      \draw[line width=\hatchlinewidth] (0,0) -- (\hatchsize,\hatchsize);
  }
}

\tikzdeclarepattern{
  name=mylines,
  parameters={
      \pgfkeysvalueof{/pgf/pattern keys/size},
      \pgfkeysvalueof{/pgf/pattern keys/angle},
      \pgfkeysvalueof{/pgf/pattern keys/line width},
  },
  bounding box={
    (0,-0.5*\pgfkeysvalueof{/pgf/pattern keys/line width}) and
    (\pgfkeysvalueof{/pgf/pattern keys/size},
0.5*\pgfkeysvalueof{/pgf/pattern keys/line width})},
  tile size={(\pgfkeysvalueof{/pgf/pattern keys/size},
\pgfkeysvalueof{/pgf/pattern keys/size})},
  tile transformation={rotate=\pgfkeysvalueof{/pgf/pattern keys/angle}},
  defaults={
    size/.initial=5pt,
    angle/.initial=45,
    line width/.initial=.4pt,
  },
  code={
      \draw [line width=\pgfkeysvalueof{/pgf/pattern keys/line width}]
        (0,0) -- (\pgfkeysvalueof{/pgf/pattern keys/size},0);
  },
}

\usetikzlibrary{arrows.meta, shadings}

\definecolor{mathematicablue}{rgb}{0.87,0.94,1}
\definecolor{mathematicadarkblue}{rgb}{0.368417, 0.506779, 0.709798}
\definecolor{mathematicaorange}{rgb}{1,0.9,0.8}
\definecolor{mathematicadarkorange}{rgb}{1,0.5,0}

\begin{document}

\author{Charlotte Bäcker}
\affiliation{Institute of Theoretical Physics, TUD Dresden University of Technology, 01062, Dresden, Germany}
\author{Krishna Palaparthy}
\affiliation{Institute of Theoretical Physics, TUD Dresden University of Technology, 01062, Dresden, Germany}
\author{Walter T. Strunz}
\affiliation{Institute of Theoretical Physics, TUD Dresden University of Technology, 01062, Dresden, Germany}

\title{Revealing the quantum nature of memory in non-Markovian dynamics on IBM Quantum}
\date{\today}

\date{\today}
\begin{abstract}
We investigate memory effects in non-Markovian dynamics on superconducting quantum processors provided by IBM Quantum. We use a collision-model approach to implement suitable single- and two-qubit dynamics with a gate-based quantum circuit.
Coupling the system of interest to an ancilla allows for a characterization of the process with respect to non-Markovian memory effects in general, as well as concerning the quantumness of that memory. We demonstrate that current noisy quantum hardware is capable of verifying quantum memory in single-qubit dynamics. We then discuss why a generalization of this dynamics to the two-qubit case cannot directly be simulated in a way that allows quantum memory to be observed. Nevertheless, we present an alternative toy example that demonstrates how quantum memory of two-qubit dynamics can be witnessed using current noisy quantum computers.
\end{abstract}

\maketitle

\section{Introduction}

The past years have witnessed tremendous advances in quantum computing, especially in terms of physical realizations and their performance \cite{philipsUniversalControlSixqubit2022, PelBaeEid2022, madsenQuantumComputationalAdvantage2022, pezmei2025, Ferracin2024efficiently, BahGuaChaLi2022, Real-Time-Faul-Tolerant-Error-Correction-2021}, shaping the era of noisy intermediate-scale quantum (NISQ) computing \cite{preskill2018}.
While the first quantum computers with very few qubits were only accessible to the experimenters themselves \cite{JonMos1998, ChuGerKub1998}, today, starting with IBM Quantum (IBMQ) in 2016, there are gate-based quantum computers with more than 100 physical qubits which can be used via cloud access by a broader audience \cite{Devitt2016}.
Since then, the IBMQ platform with access to real quantum computers as well as classical local simulations of those devices have been used to investigate different properties of quantum dynamics in various contexts, for example in the field of entanglement, steering and non-locality \cite{SeiBeyLuoStr2022b, Kuzmak2020126579, gonzalezRevisitingExperimentalTest2020, Gnatenko_2021}, quantum chemistry \cite{kandalaHardwareefficientVariationalQuantum2017, TilJonCheWosGran2020}, quantum thermodynamics \cite{ChaHuaChe2022, MelSRodSouOliSarLan2022, FelVed2020}, quantum communication \cite{dasDesignQuantumRepeater2021} and quantum machine learning  \cite{Jaderberg_2022, dasQuantumPatternRecognition2023}.

Although one major motivation to build and investigate quantum computers is their theoretical capability of solving problems which are hard to solve classically \cite{Sho1997, Gro1996}, earlier it was Feynman who advocated the idea that one should use quantum computers to simulate quantum dynamics \cite{Fey1982}. Along this line of thought, 
possible applications lie in the field of chemistry \cite{peruzzoVariationalEigenvalueSolver2014, QuantumComputationalChemistry, FaultTolerantQuantumChemistry2021,EfficientSimulationQuantumChemistry}, materials science \cite{bauerQuantumAlgorithmsQuantum2020, MaterialsScience2020}, many-body physics \cite{fausewehQuantumManybodySimulations2024} and optimization problems in general \cite{ajagekarQuantumComputingBased2020}.
For such quantum simulations it is crucial that the desired quantum features such as coherence, entanglement or quantum memory can be maintained by the quantum hardware.
However, noise and dissipation in this NISQ era lower the performance of any quantum computing device, reducing its ability to simulate specific chosen dynamics \cite{kikuchiRealizationQuantumSignal2023, reschBenchmarkingQuantumComputers2021, PanFenLiLiuLi2023} accurately.

In this work we investigate whether contemporary quantum computers are already capable of simulating non-Markovian quantum dynamics where quantum memory is verifiably involved.
Opposed to memoryless Markovian quantum dynamics, for example in terms of a Gorini-Kossakowski-Sudarshan-Lindblad master equation \cite{GorKosSud1976, Lin1975}, non-Markovian dynamics requires some notion of memory to be accounted for \cite{BreLaiPii2009, HalCreLiAnd2014, RivHuePle2014,PolRodFraPatMod2018,megierMemoryEffectsQuantum2021}. Investigating memory in open quantum dynamics experimentally is challenging \cite{liuExperimentalControlTransition2011, chiuriLinearOpticsSimulation2012}. 
Clearly, with its gate-based architecture, IBMQ provides a versatile tool to implement such quantum dynamics. More recent approaches thus also use the IBMQ quantum hardware for the purpose of investigating non-Markovian quantum dynamics \cite{MorPolKav2022, whiteDemonstrationNonMarkovianProcess2020, ChaJuCheCheKu2024}.

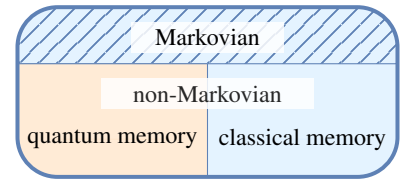
\begin{figure}
    \centering
\begin{tikzpicture}
    \draw[rounded corners=15pt] (0.5, 1.5) rectangle (5.5, 3.75);
    \draw[fill=mathematicablue, fill opacity=0.8, rounded corners=15pt]   (0.5,1.5) rectangle ++(5,2.25);
    \fill [pattern={mylines[size= 5pt,line width=0.8pt,angle=45]},  pattern color=mathematicadarkblue]
      (0.5,3) --
      ++(5,0) {[rounded corners=15] --
      ++(0,0.75) --
      ++(-5,0)} --
      cycle
      {};
    \draw[draw, color=mathematicadarkblue] 
      (0.5,3) --
      ++(5,0) {[rounded corners=15] --
      ++(0,0.75) --
      ++(-5,0)} --
      cycle
      {};
    \fill[fill=white, fill opacity=1]  
      (3,3) --
      ++(0,-1.5) {[rounded corners=15] --
      ++(-2.5, 0)} --
      ++(0, 1.5) --
      cycle
      {};
    \fill[fill=mathematicaorange, fill opacity=0.8]  
      (3,3) --
      ++(0,-1.5) {[rounded corners=15] --
      ++(-2.5, 0)} --
      ++(0, 1.5) --
      cycle
      {};
    \draw[color=mathematicadarkblue] (3, 1.5) -- ++(0, 1.5);
    \draw[color=mathematicadarkblue] (3, 3) -- ++(-2.5, 0);
    \draw[rounded corners=15pt, line width=1.5pt, color=mathematicadarkblue] (0.5, 1.5) rectangle (5.5, 3.75);
    \fill[color=white, fill opacity=0.85] (2, 3.15) rectangle (4, 3.55) node[pos=.5, color=black, opacity=1] {Markovian};
    \fill[color=white, fill opacity=0.85] (1.6, 2.4) rectangle (4.4, 2.8) node[pos=.5, color=black] {non-Markovian};
    \draw[color=mathematicaorange] (1, 2) rectangle (2.5, 2) node[pos=0.5, color=black] {quantum memory};
    \draw[color=mathematicablue] (3.5, 2) rectangle (5, 2) node[pos=0.5, color=black] {classical memory};
\end{tikzpicture}
    \caption{Classification of quantum dynamics with respect to the absence or presence of memory (Markovian and non-Markovian) as well as the type of memory (quantum and classical).}
    \label{fig:classification}
\end{figure}

It became clear that memory in non-Markovian quantum dynamics need not be quantum at all.
Sometimes memory effects can be explained with classical memory \cite{MegChrPiiStr2017, BaeBeyStr2024}. By contrast, there are also dynamics where the non-Markovianity requires genuine quantum memory \cite{BaeBeyStr2024, YuOhsNguNim2025}, see Fig. \ref{fig:classification} for a visualization of different classes of quantum dynamics. 

The quantumness of memory is a valuable resource in the quantum simulation of dynamics \cite{RosBusLia2018, BuscemiNCIR}.
Thus the aim of this paper is to investigate whether quantum simulations on NISQ computers currently provided by IBMQ are able to realize non-Markovian quantum dynamics which, in theory, rely verifiably on quantum memory.
In order to ensure that the real implementation of such dynamics maintain this crucial feature of quantum memory being involved, one has to verify its quantumness.
Classifying non-Markovian quantum dynamics by distinguishing dynamics with quantum memory from those of classical memory has become an intense field of research where multiple approaches aim at characterizing different aspects of the quantumness of memory effects 
~\cite{MilEglTarThePleSmiHue2020, GiaCos2021, BanMarHorHor2023, TarQuiMurMil2024, BaeBeyStr2024, BaeBeyStr2025:p, YuOhsNguNim2025, BaeLinStr2025:p}.
Many approaches use the powerful framework of process tensors \cite{PolRodFraPatMod2018, GiaCos2021, White2025whatcanunitary} as they carry the maximal possible information about the dynamics. Experimental advances using IBMQ have shown that obtaining a (restricted) process tensor for the classification of the memory is expensive and can so far be used to investigate single-qubit dynamics only \cite{White2025whatcanunitary}.
The aim of this article is to simulate non-Markovian dynamics and to witness the quantumness of the memory with a criterion based only on the dynamical map \cite{BaeBeyStr2024} of the dynamics.
The advantage of this map-based criterion is that it is less expensive experimentally such that not only single-qubit dynamics but also dynamics in higher dimensional quantum systems can be characterized \cite{BaeBeyStr2025:p, BaeLinStr2025:p}.

Note that the authors of Ref.~\cite{ChaJuCheCheKu2024} also use the notion \emph{quantum memory} but in the sense of \emph{quantum} dynamics with memory contrasting \emph{classical} dynamics with memory. In fact, the dynamics investigated there as well as one of the first tunable non-Markovian dynamics experimentally implemented and controlled from Ref.~\cite{liuExperimentalControlTransition2011} is of the random unitary type and thus it has been shown to not require quantum memory in the strict sense we will use here, see Ref.~\cite{BaeBeyStr2024}.

Using the map-based definition of classical and quantum memory we investigate models which have been shown to require quantum memory in theory. These dynamics are simulated on the quantum computer using a collision model. Due to noise on the NISQ computer, the real quantum simulation will deviate from theory and our aim is to investigate whether the noise is too strong for the quantum nature of the memory to be witnessed or whether the noisy quantum simulation still allows for a characterization of the memory as verifiably quantum. As a third pillar, next to ideal theory and quantum simulation, we show
results of the (classical) local simulation of the noisy quantum dynamics based on the noise models provided by the IBMQ platform. 

This paper is structured as follows. First, we introduce the concept of quantum memory in non-Markovian dynamics in Sec.~\ref{sec:quantum_memory}. In Sec.~\ref{sec:single_qubit} we investigate a model of single-qubit dynamics, its quantum simulation on IBMQ hardware, and its local simulation using the IBMQ noise models. First, we discuss its non-Markovianity with respect to different criteria followed by witnessing the quantumness of the memory. 
A similar approach as for the single-qubit dynamics is then applied to a two-qubit dynamics in Sec.~\ref{sec:two_qubit}. Finally, we provide our conclusions and an outlook in Sec.~\ref{sec:conclusions}.

\section{Non-Markovianity and quantum memory}
\label{sec:quantum_memory}

While Markovianity (and thus non-Markovianity) is well defined for a process in classical physics or mathematics, this does not hold for the quantum case \cite{VacSmiLaiPiiBre2011}.
Indeed, there are multiple approaches to define non-Markovianity in quantum physics. Some, for example, are based on monotonicity of distance measures, correlations or geometric quantities \cite{RivHuePle2014, BreLaiPii2009, RivHuePle2010, LorPlaPat2013}. 
Non-Markovian dynamics arises from some form of memory -- the future is not only determined by the presence, but by the history leading to the present state.
However, the appearance of quantum non-Markovianity does not by itself imply that the underlying memory effects are of truly quantum origin. We may well have non-Markovian quantum dynamics where the underlying memory may entirely be described by classical data. In order to distinguish the physical nature of the required memory, several concepts and definitions of classical or quantum memory in various frameworks have been introduced, which differ mainly in the amount of information available to an experimenter \cite{BaeBeyStr2024, GiaCos2021, TarQuiMurMil2024, BanMarHorHor2023, VieKuBud2024}. We will use the minimum-information map-based definition of classical and quantum memory provided in Ref.~\cite{BaeBeyStr2024}. This is based on a discrete   \emph{dynamics} \(\Dyn\) of an open quantum system \(\Sys\) which is a finite, ordered set of completely positive trace-preserving maps (CPT maps)  \(\Dyn = (\cpt_{t_1},\cpt_{t_2},\ldots\cpt_{t_N})\) with \(t_1<t_2< \ldots <t_N\) emerging from a dynamical map $\cpt_t$, mapping the system state from the initial time \(t_0\) to time \(t>t_0\). The advantage of considering maps at distinct times instead of process tensors lies in the comparably low experimental effort to obtain the relevant information \cite{white2022Non, White2025whatcanunitary}.

Here, we focus on a two-time dynamics  \(\Dyn = (\cpt_{t_1},\cpt_{t_2})\) and say that it
is \emph{realizable with classical memory} if there exists a set of Kraus operators $\{ K_i\}$
and a set of CPT maps $\Phi_i$ such that
\begin{align}\label{eq:map_def}
        \cpt_{t_1}[\rho] = \sum_i K_i \rho K_i^\dagger, && \cpt_{t_2}[\rho] = \sum_i \Phi_i[K_i \rho K_i^\dagger].
\end{align}
Otherwise the dynamics is said to require quantum memory.
Dynamics with classical memory can be realized in two steps. First, at $t=t_1$, a measurement on the system is performed such that on average the map $\cpt_{t_1}$ is realized. Now, conditioned on the outcome $i$ of this measurement, a CPT map $\Phi_i$ is applied. This yields the overall map $\cpt_{t_2}$ when averaged over all measurement outcomes $i$. Crucially, the measurement outcome $i$ can be stored in classical memory, and the subsequent evolution given by the conditioned CPT map $\Phi_i$ can be realized with a new, uncorrelated environment. 
Related constructions can be found in Refs.~\cite{LiHalWis2018, TarQuiMurMil2024}.

Given the two maps \(\Dyn = (\cpt_{t_1},\cpt_{t_2})\), there is no straightforward way to find out whether there exist suitable combinations of Kraus operators $K_i$ and maps $\Phi_i$ such that the memory can be modeled classically as in Eqs.~\eqref{eq:map_def}. However, in Refs.~\cite{BaeBeyStr2024, BaeBeyStr2025:p, BaeLinStr2025:p,  YuOhsNguNim2025, beyer2025onesidedwitnessquantumnessgravitational}  it has been shown that there are sufficient criteria to rule out the possibility that the dynamics can be written in terms of classical memory (as in Eqs.~\eqref{eq:map_def}), and thus to witness the quantumness of the memory.

It turns out very useful to extend the two CPT maps \(\cpt_{t_1}\) and \(\cpt_{t_2}\)  on a system $\Sys$ to an initially entangled joint system-ancilla state given by $\rhoSAzero$. The time-evolved joint state at time $t$ then takes the form
\begin{align}
\label{eq:initial_sys_anc_state}
    \rhoSAt = (\cpt_{t}\otimes \id_\Anc)[\rhoSAzero].
\end{align}
For the maximally entangled $\rhoSAzero =\ket{\Phi_+}\bra{\Phi_+}$ with $\ket{\Phi_+}\sim \sum_j\ket{jj}$, $\rhoSAt$ is nothing but the Choi-Jamiołkowski state corresponding to the map $\cpt_{t}$.
Now, as shown in Refs.~\cite{BaeBeyStr2024, BaeBeyStr2025:p}, if we observe that the joint states  \(\rhoSAone\) and \(\rhoSAtwo\) at times \(t_1,t_2\) satisfy
    \begin{align}
    \label{eq:theorem}
        \conc^\sharp\left[\rhoSAone\right] < \conc\left[\rhoSAtwo\right],
    \end{align}
where $\conc$ is the concurrence of formation \cite{Wootters1998} and $\conc^\sharp$ is the concurrence of assistance \cite{DiVFucMabSmoThaUhl1999, LauVerEnk2002}, the memory has to be quantum and the dynamics cannot be realized with classical memory~\cite{BaeBeyStr2024, BaeBeyStr2025:p}. 
We will apply this criterion to single- and two-qubit dynamics implemented on an IBMQ NISQ computer. Note that for a single qubit system, when $\rhoSA$ is a two-qubit state, there are closed-form expressions for both concurrences.

\section{Single-qubit dynamics}
\label{sec:single_qubit}

In order to witness quantum memory in non-Markovian dynamics on the hardware provided by IBMQ we will first consider dynamics on the smallest possible system, namely single-qubit dynamics.
The qubit Hilbert space is spanned by the states $\ket{0}$ and $\ket{1}$.

\subsection{The Model}
Starting point is a physically motivated non-Markovian amplitude damping process in continuous time \cite{breuerTheoryOpenQuantum2007,kretschmerCollisionModelNonMarkovian2016,garrawayNonperturbativeDecayAtomic1997,diosiNonMarkovianQuantumState1998}.
In the zero-temperature case the dynamics is governed in terms of the master equation
\begin{align}
\label{eq:nMadthLindblad}
\dot \rho=\mathcal{L}_t\left[\rho\right] = \frac{\gamma_-(t)}{2} \left(\left[\sigma_- \rho, \sigma_+\right] + \left[\sigma_- ,\rho  \sigma_+\right]\right),
\end{align}
where $\sigma_- = \ket{0}\bra{1} = \sigma_+^\dagger$ and $\gamma_-(t)=\tan(t)$ changes sign periodically. Thus, the dynamics is non-Markovian according to all the common definitions of quantum non-Markovianity \cite{HalCreLiAnd2014, RivHuePle2010, BreLaiPii2009, LorPlaPat2013}.

Furthermore, it has been shown that the memory effects necessary to realize this dynamics are caused by memory which is truly quantum according to the definition given in Eqs.~\eqref{eq:map_def}, using the criterion in Eq.~\eqref{eq:theorem}, see Ref.~\cite{BaeBeyStr2024} for details\footnote{Note that compared to the model considered in Ref.~\cite{BaeBeyStr2024} where quantum memory was discussed, here we set $\gamma=0=\eta$ and $g_0=1$, such that there is no additional damping on the environmental qubit and we obtain the ideal zero-temperature case showing full revival to the initial state of the example discussed in Ref.~\cite{BaeBeyStr2024}.}.
The image of the Bloch sphere for a dynamics described by Eq.~\eqref{eq:nMadthLindblad} periodically contracts to a point at the north pole $\ket{0}$ and then regrows to the full initial sphere. 
\begin{table*}[]
\begin{ruledtabular}
    \centering
    \begin{tabular}{l|ccccc}
    Theory &\raisebox{-0.5\totalheight}{\includegraphics[width=0.17\linewidth]{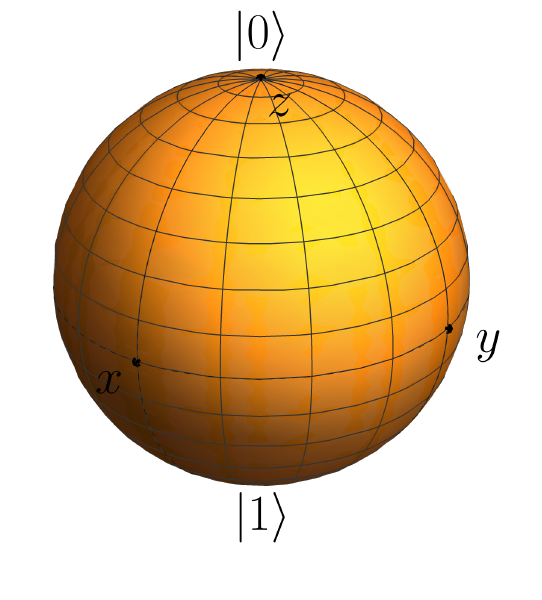}} & \raisebox{-0.5\totalheight}{\includegraphics[width=0.17\linewidth]{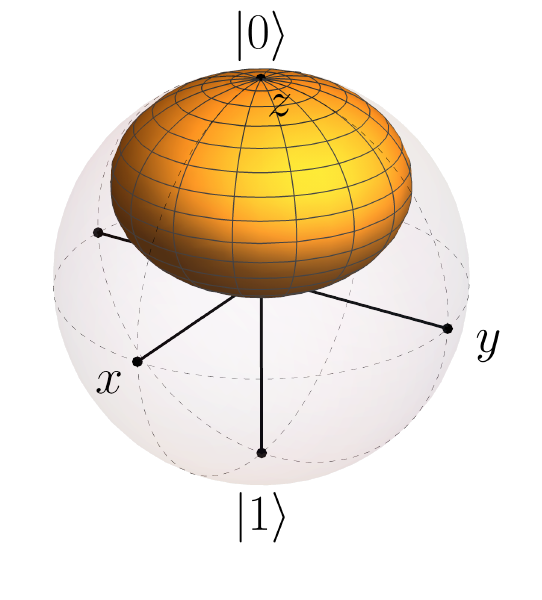}}& \raisebox{-0.5\totalheight}{\includegraphics[width=0.17\linewidth]{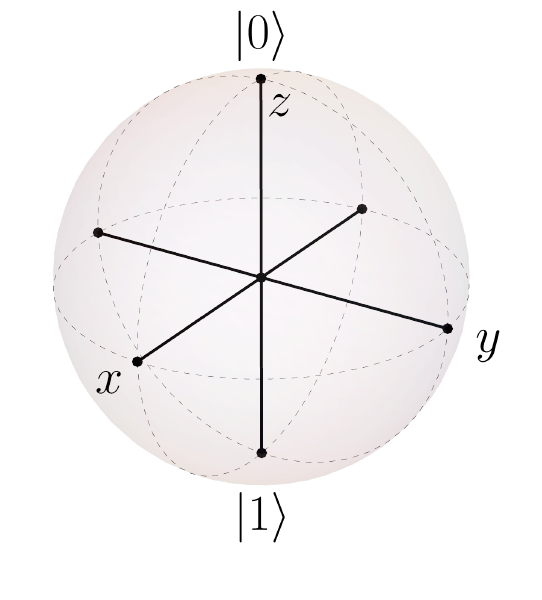}}& \raisebox{-0.5\totalheight}{\includegraphics[width=0.17\linewidth]{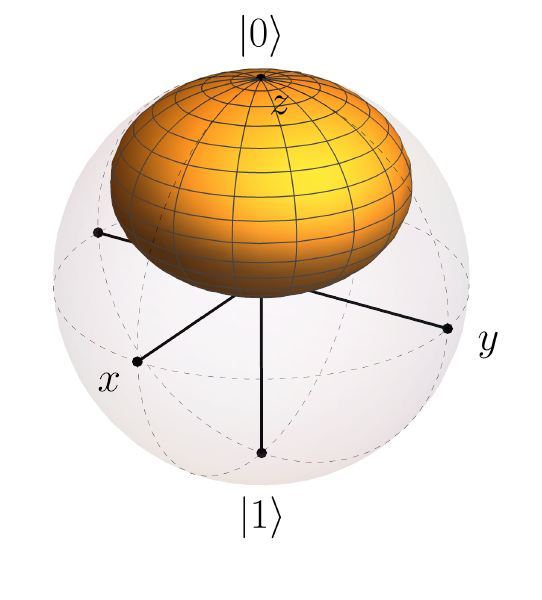}}& \raisebox{-0.5\totalheight}{\includegraphics[width=0.17\linewidth]{blochplot_analytical_0.pdf}}\\
    \hline
   \makecell{Quantum\\ simulation} &\raisebox{-0.5\totalheight}{\includegraphics[width=0.17\linewidth]{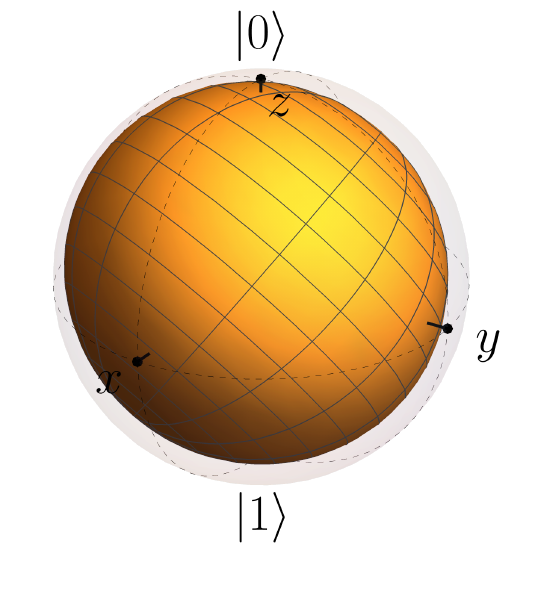}} &
   \raisebox{-0.5\totalheight}{\includegraphics[width=0.17\linewidth]{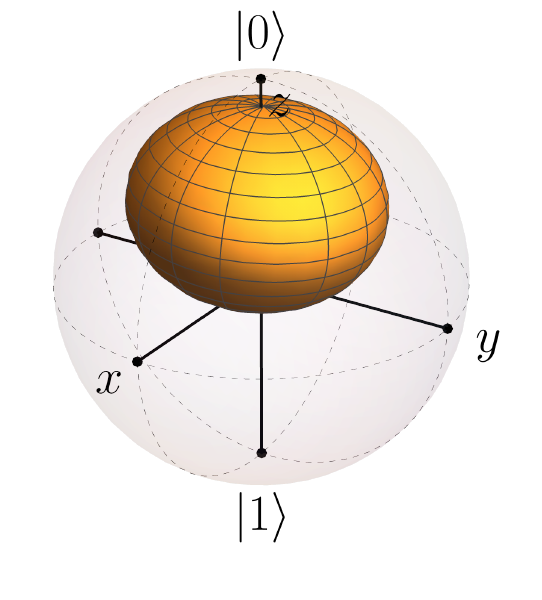}}&
   \raisebox{-0.5\totalheight}{\includegraphics[width=0.17\linewidth]{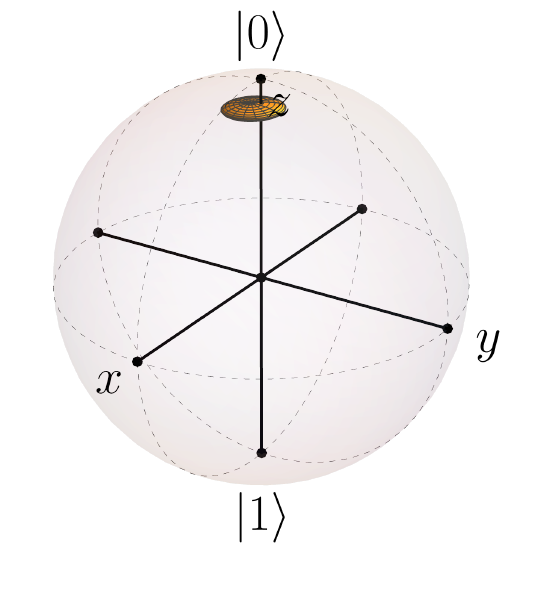}}&
   \raisebox{-0.5\totalheight}{\includegraphics[width=0.17\linewidth]{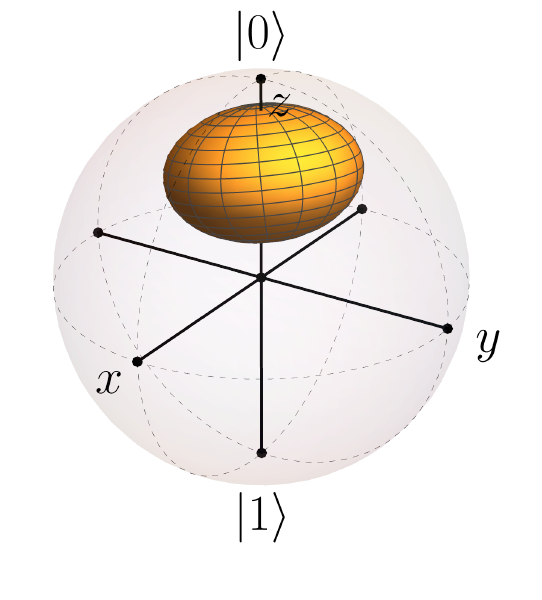}}&
   \raisebox{-0.5\totalheight}{\includegraphics[width=0.17\linewidth]{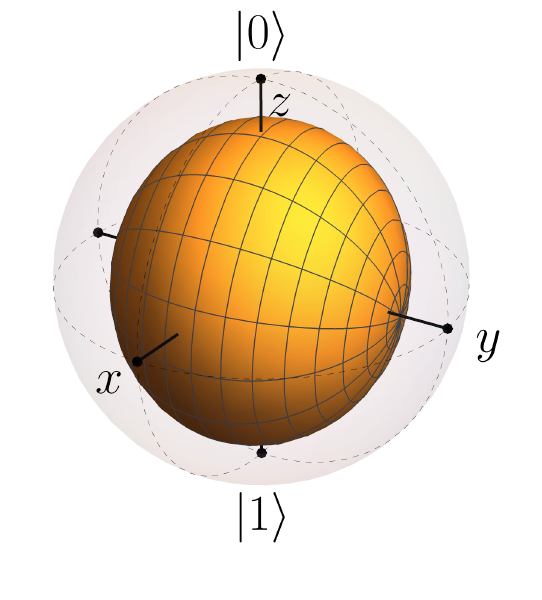}}\\
    \hline
   Collisions $n$& $0$ & $1$ & $2$ & $3$ & $4$\\
   \hline
   Fidelity  & 0.86 & 0.88 &0.93 & 0.79 & 0.57\\
    \end{tabular}
    \caption{Evolution of the hull of the Bloch ball of the system under the dynamics described by Eq.~\eqref{eq:coll_mod} for the choice of $\gdt=\pi/4$ for up to four collisions. Upper: Theoretical model without additional noise from quantum hardware. Lower: Reconstructed image of the Bloch sphere resulting from the quantum simulation on \texttt{ibm\_sherbrooke} according to the circuit in Fig.~\ref{fig:circuitmodel}. In the last line the fidelity between the theoretical Choi-Jamiołkowski state and the one obtained from the implementation on IBM Quantum rounded to two decimals is shown.
    It was assumed that the noise on the ancilla is low enough such that the time-evolved system-ancilla state could be identified with the Choi-Jamiołkowski state of the dynamical map on the system.}
    \label{tab:tabbloch}
\end{ruledtabular}
\end{table*}

Witnessing the quantumness of the memory requires the investigation of the map at two different times.
Choosing $t_1=\pi/2$ and $t_2=\pi$ we find that for the theoretical model $\conc^\sharp(\rhoSA_{t_1})=0 \leq 1 = \conc(\rhoSA_{t_2})$ and hence according to Eq.~\eqref{eq:theorem} realizing this dynamics strictly requires quantum memory. Since this is the maximal possible difference between $\conc^\sharp(\rhoSA_{t_1})$ and $\conc(\rhoSA_{t_2})$, the dynamics described by the master equation~\eqref{eq:nMadthLindblad} is a suitable candidate for the detection of quantum memory in non-Markovian dynamics.
In order to implement such time-continuous dynamics in a time-discrete way suitable for the IBMQ hardware using the Qiskit SDK~\cite{qiskit2024}, we will transform the time-continuous dynamics into a so-called collision model~\cite{ZimSteBuz2005, ciccarello2022}.
The idea is to model the environment in terms of many sub-environments which sequentially interact with the system via short-time unitaries, so-called \emph{collisions}. 
The local dynamics is thus a CPT map by construction and can be Markovian or non-Markovian, depending on whether there is interaction between the sub-environments feeding forward information concerning the history of the process, or not \cite{KreLuoStr2016}.
In the Markovian case, each of the collisions happens with a fresh, independent sub-environment such that no correlations are preserved. In the case of non-Markovian dynamics, the aim is to preserve some system-environment correlations. 

We would like the global system-environment-state to retain as much information as possible and thus only choose one qubit representing our environment to model the single-qubit non-Markovian amplitude damping from Eq.~\eqref{eq:nMadthLindblad}. This qubit then repeatedly collides with the system qubit, ensuring that the maximum information concerning the history of the dynamics can, in principle, influence the present system qubit state.
As the global unitary operation $\ug$ acting on system and environment we choose
\begin{align}
    \label{eq:ug}
\ug &= \e^{\i \gdt (\sigma_- \otimes \sigma_+ + \sigma_+ \otimes \sigma_-)}\notag\\
    &= \begin{pmatrix}
        1 & 0 & 0 & 0\\
        0 & \cos(\gdt) & -\i \sin(\gdt)& 0\\
        0 &-\i \sin(\gdt) & \cos(\gdt) & 0\\
        0 & 0 & 0 & 1
    \end{pmatrix},
\end{align}
where $\gdt$ is a measure of the strength of the collision.

The time-discrete local dynamics $\Dyn=\left(\cpt_0, \cpt_1, ..., \cpt_N\right)$ of Eq.~\eqref{eq:nMadthLindblad} then consists of maps taking the form 
\begin{align}
    \label{eq:coll_mod}
    \cpt_n\left[\rho_{\Sys}\right] = \mathrm{tr}_\Env \left[\ug^n \left(\rho_{\Sys} \otimes \rho_{\Env} \right)\left(\ug^\dagger\right)^n\right].
\end{align}
where $\rho_\Sys \otimes \rho_\Env$ is an initially uncorrelated system-environment state.
A graphical depiction of the dynamics in terms of its action on the Bloch sphere is shown in the upper row in Tab.~\ref{tab:tabbloch}.
In the time-continuous limit $n\rightarrow \infty$ with $\gdt = gt/n \to 0$ the local map on the system $\Sys$ is given by $\cpt_n \rightarrow \cpt_t$ corresponding exactly to the dynamics described by the master equation Eq.~\eqref{eq:nMadthLindblad}, this is shown in detail in App.~\ref{sec:time_continuous_limit}.

Knowing the maps from the dynamics $\Dyn = (\cpt_1, \cpt_2, ..., \cpt_N)$ is sufficient to verify the quantumness of the memory according to Eq.~\eqref{eq:theorem}.
However, since we would like to implement the dynamics on real hardware, there is a more suitable approach which does not require channel tomography of the map.

\subsection{Implementation on IBMQ}
\label{sec-implementation-ibmq}
\begin{figure}
    \centering
    \begin{quantikz}[row sep={0.8cm,between origins}, slice style=blue]
        \lstick{$\rho_\Anc$} & \gate{H} & \ctrl{1} & \slice{$t_1$} & \slice{$t_2$} & \slice{$t_3$} & \rstick[2]{$\rho_{\Sys \Anc}^{(3)}$} \\
        \lstick{$\rho_\Sys$} & & \targ{}&\gate[2]{U} & \gate[2]{U} & \gate[2]{U}  &\\
        \lstick{$\rho_\Env$} & & & & & &
    \end{quantikz}

    \caption{Time-discrete implementation of  the dynamics described by Eq.~\eqref{eq:nMadthLindblad} according to Eq.~\eqref{eq:coll_mod}. All qubits are by default initialized in the $\ket{0}$ state. First, system and ancilla are prepared in a maximally entangled state and the environment is left in the $\ket{0}$-state. Then the unitary $\ug$ is applied to the system-environment state sequentially up to $N$ times, here we depict the circuit for $N=3$ collisions. Finally, quantum state tomography is performed on the system-ancilla state $\rho_{\Sys \Anc}^{(3)}$. The actual implementation of the Bell state as well as of the unitary $\ug$ in terms of the fundamental basis gates of the quantum hardware can be found in App.~\ref{sec:app-implementation}.}
    \label{fig:circuitmodel}
\end{figure}
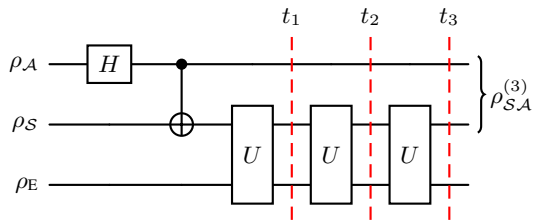

In order to witness quantum memory, instead of extracting the map and computing its related Choi-Jamiołkowski state one can also directly monitor the dynamics of the system coupled to an ancilla, i.e. we aim at a physical realization of the Choi-Jamiołkowski-isomorphism, see Eq.~\eqref{eq:theorem}.
Thus, we initialize system and ancilla in a maximally entangled Bell state and leave the ancilla untouched afterwards. The environmental qubit is initially in the $\ket{0}$-state and then repeatedly interacts with the system qubit. For a graphical representation of this quantum circuit see Fig.~\ref{fig:circuitmodel}.

We implement the collision model for all numbers $n\in\left(0, ..., N=10\right)$ of collisions on the IBMQ computing resource \texttt{ibm\_sherbrooke}, which is one of the IBMQ Eagle processors. 
After each of the $N+1$ circuits we perform quantum state tomography of the system-ancilla state $\rho_{\Sys \Anc}^{(n)}$, which requires nine different tomography settings in total. For each of these settings we run 4096 shots which yields in total 36.864 runs per circuit.
Assuming that the ancilla is isolated from the environment, the system-ancilla states $\rho_{\Sys \Anc}^{(n)}$ can be used to reconstruct the quantum map with the help of the Choi-Jamiołkowski isomorphism. The image of the Bloch sphere under those maps obtained from the quantum simulation on \texttt{ibm\_sherbrooke} is depicted in the second line in Tab.~\ref{tab:tabbloch}, where a deviation from the theoretical expectation in the first line can be observed.
We will now first have a look on the non-Markovianity of the resulting dynamics and afterwards discuss the quantumness of the memory.

\subsubsection{Non-Markovianity}
\label{sec:criteria}

The crucial times in the theoretical model are $t_1$ corresponding to two collisions and $t_2$ corresponding to four collisions. We will now discuss the non-Markovianity of the dynamics $\Dyn=(\cpt_{t_1}, \cpt_{t_2})$ with respect to different proposed criteria.

A widely used criterion for non-Markovian quantum dynamics is an increase in the entanglement with an ancilla \cite{RivHuePle2010}. Since we already implemented the dynamics with such an additional ancilla, we can directly evaluate this measure for our dynamics.
For both, theory (subscript $\text{t}$) and quantum simulation (subscript $\text{qs}$) we observe $\conc_\text{t}(t_1)=\conc_\text{qs}(t_1)\approx 0$. At time $t_2$ we find $\conc_\text{t}(t_2)=1$ and $\conc_\text{qs}(t_2)\approx 0.62$, respectively. Thus, in both cases it is evident that there is an increase in concurrence and hence entanglement with the ancilla reflecting the non-Markovian nature of the dynamics. This is also a sufficient criterion for the dynamics  $\Dyn = (\cpt_{t_1}, \cpt_{t_2})$ to be CP-indivisible~\cite{RivHuePle2010}.

However, dynamics which is CP-indivisible can still be P-divisible and thus P-indivisibility is regarded as a stricter definition of non-Markovianity~\cite{BreLaiPii2009}.
A sufficient criterion for P-indivisibility is an increase of the trace distance between two initial states under the dynamics.
Since we used the maximally entangled state as our initial state, we have direct access to the noisy Choi-Jamiołkowski state and thus also to the noisy map. Optimizing over the whole set of initial states, we find that the maximum increase of the trace distance for those two times ${\Delta(d_\text{tr}(t_2), d_\text{tr}(t_1)})_\text{qs} \approx 1.72$ in the quantum simulation and ${\Delta(d_\text{tr}(t_2), d_\text{tr}(t_1)})_\text{t} = 2$ from theory.
The two initial states considered in the experimental case (in the theoretical case any two points which are opposite on the Bloch sphere reach an increase from distance 0 to distance 2) are $\vec{r}_a = (-0.25, -0.92, 0.29)$ and  $\vec{r}_b = (0.25, 0.92, -0.29)$. 
An even stricter criterion for non-Markovianity requires also the volume of the Bloch sphere $V_\mathrm{Bloch}$ to increase under the dynamics \cite{LorPlaPat2013}. The volume of the Bloch sphere can also be obtained via the noisy map following Refs.~\cite{MilJevJenWisRud2014, JevPusJenRud2014}. In theory the Volume is $V_\mathrm{Bloch, t}(t_1)=0$ at $t_1$ and $V_\mathrm{Bloch, t}(t_2)=V_0$ at $t_2$ while in the quantum simulation $V_\mathrm{Bloch, qs}(t_1)\approx 0.005 V_0$ and $V_\mathrm{Bloch, qs}(t_2)\approx 0.41 V_0$ with $V_0=\frac{4}{3}\pi$. Although the increase in volume obtained from the quantum simulation is not even half of the increase one would expect from theory, it is evident that the dynamics realized on the quantum computer is also non-Markovian with respect to this very strict criterion.

To summarize, the implemented dynamics clearly shows memory effects and is non-Markovian with respect to weaker and stricter criteria and definitions of quantum non-Markovianity although the effects are weaker than expected from theory. We will now investigate whether the observed memory effects nevertheless genuinely require quantum memory or could in principle be explained with classical memory according to our definition from Eqs.~\eqref{eq:map_def}.

\subsubsection{Quantum memory}

\begin{figure}
    \centering
    \includegraphics[width=\linewidth]{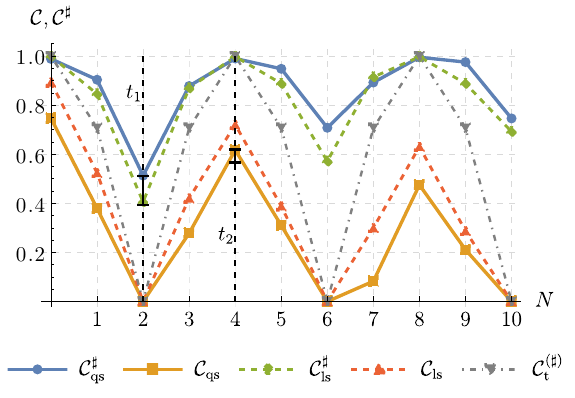}
    \caption{Concurrence of formation $\conc$ and concurrence of assistance $\conc^\sharp$ of the system-ancilla state under the system dynamics realized via the collision model as described in Eq.~\eqref{eq:coll_mod} where we chose $\gdt=\pi/4$. The quantum simulation (subscript qs) was run on 2025/05/28 on the IBM Quantum computer \texttt{ibm\_sherbrooke} with 4096 shots for each of the nine tomography settings in each circuit. The dashed curves corresponds to a local simulation (subscript ls) on \texttt{fake\_sherbrooke} and the dot-dashed curve represents the analytical results without any additional noise or noise models (subscript t). 
    The concurrence of assistance at $t_1$ (2 collisions) is lower than the concurrence of formation at $t_2$ (4 collisions) implying that the memory is necessarily quantum according to Eq.~\eqref{eq:theorem}. The error bars shown for those two points reflect the fluctuating performance of \texttt{ibm\_sherbrooke} and have been obtained by collecting statistics for the corresponding circuits at different times within a period of two weeks. The average of several runs of $\conc^\sharp$ and $\conc$ for two and four collisions lies in the middle of the error bars while the curves themselves represent the average of only one example run per circuit under identical conditions.}
    \label{fig:basicquantummemory}
\end{figure}

According to Eq.~\eqref{eq:theorem} we can use the reconstructed $N+1$ system-ancilla states $\rho_{\Sys \Anc}^{(n)}$ to verify the quantumness of the memory of the dynamics. For this purpose we compute the concurrence of formation and the concurrence of assistance of the joint time-evolved system-ancilla state. The results from quantum and local simulation as well as the theoretical predictions are depicted in Fig.~\ref{fig:basicquantummemory}.
Note that \emph{quantum simulation} here refers to the actual quantum computer \texttt{ibm\_sherbrooke}, whereas \emph{local simulation} refers to computations on \texttt{fake\_sherbrooke}, a classical emulator based on realistic noise models. The theoretical values stem from calculations of the ideal noiseless dynamics in terms of the collision model. In said figure also an estimate of the error due to statistical fluctuations of noise and readout error can be found.
Comparing the behavior of the concurrences with respect to time can be used to witness quantum memory according to Eq.~\eqref{eq:theorem}.
As we can see in Fig~\ref{fig:basicquantummemory}, the concurrence of assistance at $t_1$ corresponding to two collisions is below the concurrence of formation at $t_2$ two collisions later,
\begin{align}
    \label{eq:conc-result}
    \conc^\sharp(t_1)=0.51 < 0.62 = \conc(t_2).
\end{align}
Hence, the quantumness of the memory provided by the environmental qubit in the implementation Fig.~\ref{fig:circuitmodel} can be verified via Eq.~\ref{eq:theorem}. 
This is remarkable because as one can see in Tab.~\ref{tab:tabbloch}, the fidelity between the theoretical and observed Choi-Jamiołkowski state at $t=t_1$ is $F_{t_1}(\rhoSA_\text{t}, \rhoSA_\text{qs})\approx 0.93$, yet it is only  $F_{t_2}(\rhoSA_\text{t}, \rhoSA_\text{qs})\approx0.57$  at $t=t_2$. Still, the quantum nature of the memory in this experimental non-Markovian process on a quantum computer can still be witnessed. We note in passing that
this result has also been confirmed qualitatively on the quantum computers \texttt{ibm\_kyiv} and \texttt{ibm\_brisbane}, which are also instances of the Eagle processor.

Note that the data from the local simulation depicted in Fig.~\ref{fig:basicquantummemory} would account for an additional combination of times which can be used to witness quantum memory. Taking the value of the concurrence of assistance after six collisions and the concurrence of formation after eight collisions, it becomes clear that the performance of the quantum simulation on the actual NISQ computer stays behind the performance of the classical local simulation on \texttt{fake\_sherbrooke}.

Note that it is also possible to investigate the case $\gdt = \pi/2$ such that $t_1$ corresponds to one collision and $t_2$ to two collisions. Since this reduces the number of gates necessary to implement the dynamics and thus also the noise, the fidelities between expected and observed states are higher, at $t=t_1$ it is approximately $F_{t_1}(\rhoSA_\text{t}, \rhoSA_\text{qs})\approx 0.96$, and even at $t=t_2$ $F_{t_2}(\rhoSA_\text{t}, \rhoSA_\text{qs})\approx0.81$. This leads to a more substantial difference of the concurrences than observed in Eq.~\eqref{eq:conc-result}, for the case $\gdt = \pi/2$ we find $\conc^\sharp(t_1)=0.39 < 0.77 = \conc(t_2)$.

\section{Two-qubit dynamics}
\label{sec:two_qubit}

In this section we show how the map-based criterion can also be applied to characterize memory in two-qubit dynamics.

\subsection{Generalization of the single-qubit dynamics}
\label{sec:two_qubit_physics}

The two-qubit dynamics we will investigate first is motivated by extending the above single-qubit dynamics to two system qubits. 
In order to minimize unnecessary noise from the quantum hardware, we aim at using the least possible number of non-system qubits. Thus, the environment is chosen to be a single qubit such that we can consider a three-qubit unitary $\ug =  \e^{\i H \dt}$ with
\begin{align}
    \label{eq:h_u_2qub}
    H =\ &\sigma_- \otimes \sigma_+ \otimes \id + \sigma_+ \otimes \sigma_- \otimes \id \notag \\  + &\id \otimes \sigma_- \otimes \sigma_+ + \id \otimes \sigma_+ \otimes \sigma_-.
\end{align}

Different to the single-qubit dynamics we do not start in a Bell state of system and ancilla, but we choose our initial state to be a system-ancilla state where a single ancilla is entangled with one half of the system 
\begin{align}
    \label{eq:bisep_initial}
    \ket{\psi_{\Anc \Sys_1 \Sys_2}} = \frac{1}{\sqrt{2}} \left(\ket{00} + \ket{11} \right)\ket{0}.
\end{align}
The full circuit is thus identical to the one depicted in Fig.~\ref{fig:circuitmodel}, apart from an additional system-qubit which is inserted between the other system-qubit and the environment and which is left untouched until the first collision with the unitary $\ug$. As before, after the desired number of collisions, quantum state tomography is performed on the two system qubits and the ancilla qubit.

Since the system-ancilla state is not a two-qubit state as in the single-qubit case, the criterion in Eq.~\eqref{eq:theorem} cannot be used to witness quantum memory because there is no closed-form expression for the higher-dimensional concurrence of formation $\conc$ or concurrence of assistance $\conc^\sharp$ \cite{RunBuzCavHilMil2001, Uhl2000}.
However, considering a suitable upper bound $\conc^\sharp_>$  for $\conc^\sharp$ \cite{LiFeiAlbLiu2009} and lower bounds $\conc_<$  for $\conc$ such as those in Refs.~\cite{CheAlbFei2005, MinBuc2007, WanFei2025, LuSu2025}, one arrives at the condition
\begin{align}
    \label{eq:theo_iconc}
    \conc^\sharp_>(t_1) < \conc_<(t_2)
\end{align}
which is still able to witness the necessity of quantum memory to realize the dynamics \cite{BaeLinStr2025:p}.
Here we use
\begin{align}
    \conc^\sharp_> = \sqrt{2 \left(1 - \tr \left(\tr_{\Anc}(\rho_{\Sys \Anc})^2\right)\right)}, 
\end{align}
as upper bound for the concurrence of assistance, and \cite{CheAlbFei2005}
\begin{align}
    \conc_< =\tilde{m}\max\big\{ &(||(\rho_{\Sys \Anc})^{T_\Sys}||-1), (||(\rho_{\Sys \Anc})^{T_\Anc}||-1)\big\}
\end{align}
as lower bound for concurrence. Here, $\tilde{m}=\sqrt{\frac{2}{m(m-1)}}$, with $m$ being the dimension of the smaller one of the Hilbert spaces of the two parties, and $T_{\Sys / \Anc}$ is the partial transpose with respect to system $\Sys$ or ancilla $\Anc$.

\begin{figure}
    \centering
    \includegraphics[width=\linewidth]{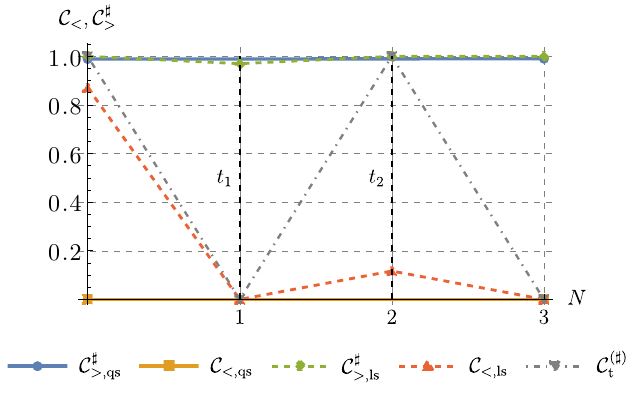}
    \caption{Lower bound of concurrence of assistance $\conc_<^\sharp$ and upper bound of concurrence of formation $\conc_>$ of the three-qubit system-ancilla state under the system dynamics described by the Hamiltonian from Eq.~\eqref{eq:h_u_2qub}. The quantum simulation was executed on 2025/05/19 on \texttt{ibm\_sherbrooke} with 4096 shots for each of the 27 tomography settings in each circuit.}
    \label{fig:2qubits_noqm}
\end{figure}

As displayed in Fig.~\ref{fig:2qubits_noqm}, in theory, the dynamics obtained from this model clearly shows that quantum memory is required to realize the dynamics.
However, already the local simulation on \texttt{fake\_sherbrooke} only shows a minor increase in $\conc_<$ at $t_2$ and an even smaller decrease of $\conc_>^\sharp$ at $t_1$ such that no quantum memory can be witnessed on the classically simulated quantum computer. The quantum simulation on real hardware \texttt{ibm\_sherbrooke} does perform even worse, there the decrease and increase are barely noticeable.
The reason for this incapability of showing signs of quantum memory is most likely due to the complex implementation of the unitary Eq.~\eqref{eq:h_u_2qub}. After the internal process of transpiling one finds that already one single collision with this unitary  requires more than 500 quantum gates according to the default transpiling operation for \texttt{ibm\_sherbrooke}. Executing this sequence of gates most likely takes longer than the decoherence times $\mathrm{T}1 \approx 280 \unit{\micro s}$ and $\mathrm{T}2 \approx 180\unit{\micro s}$, which is the time the quantum computer is able to uphold highly coherent and thus also entangled states. This means that noise effects play a tremendous role and the quantum states become more classical.
This is also reflected by the almost constant value of $\conc^\sharp$, which tells us that the dynamics is close to random unitary dynamics~\cite{DiVFucMabSmoThaUhl1999} and thus almost no quantum memory is present in the dynamics \cite{BaeBeyStr2024}. Note that the local simulation at least shows an increase of entanglement between system and ancilla which means the dynamics is non-Markovian with respect to that definition \cite{RivHuePle2010}. This, however, cannot be seen on the real quantum computer, there the entanglement remains zero once it reaches that value.

\subsection{Toy Model}
\label{sec:two_qubits_toy_model}
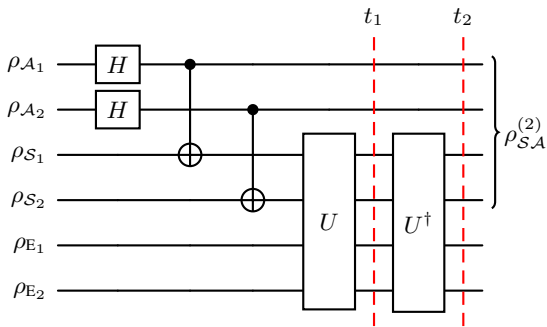
\begin{figure}
    \centering
    \begin{quantikz}[row sep={0.6cm,between origins}]
        \lstick{$\rho_{\Anc_1}$} & \gate{H} & \ctrl{2} & & \slice{$t_1$} & \slice{$t_2$} & \rstick[4]{$\rho_{\Sys \Anc}^{(2)}$} \\
        \lstick{$\rho_{\Anc_2}$} & \gate{H} & & \ctrl{2} & & &  \\
        \lstick{$\rho_{\Sys_1}$} & & \targ{}&& \gate[4]{U}\vphantom{\dagger}\hphantom{\dagger} & \gate[4]{U^\dagger} &\\
        \lstick{$\rho_{\Sys_2}$} & & & \targ{}& & &\\
        \lstick{$\rho_{\Env_1}$} & & & & & &\\
        \lstick{$\rho_{\Env_2}$} & & & & & &
    \end{quantikz}

    \caption{Quantum circuit implemented for the purpose of witnessing quantum memory in a two-qubit dynamics. The system as well as the environment consist of two qubits each and system and ancilla are prepared in a maximally entangled state. We run this circuit both, until time $t_1$ and until time $t_2$ on \texttt{ibm\_sherbrooke} on 2025/06/27 for 4096 shots for each of the 81 tomography settings and the two additional settings for readout-error mitigation. At time $t_2$ this model in theory returns to the initially entangled system-ancilla state by construction.}
    \label{fig:2qubits_circuit}
\end{figure}

Since the physically motivated example in the previous subsection required a complex transpilation to elementary gates of the chosen hardware, we aim at reducing the actual number of gates that has to be executed while maintaining quantum memory in order to create an ideal toy model.
Furthermore, we loosen the idea that the dynamics can be understood within the context of a collision model of a physical dynamics. The circuit we consider is depicted in Fig.~\ref{fig:2qubits_circuit}.

One might be tempted to choose the unitary $U$ to be a full SWAP of system and environment. In this way -- considering fundamental laws of quantum mechanics -- a dynamics is obtained, which by construction can trivially only be realized using quantum memory. The implementation of a SWAP gate on the real quantum computer, however, is comparably complex to the physically motivated example of the previous section. The amount of single- and two-qubit gates is not drastically reduced.
A quantum simulation on \texttt{ibm\_sherbrooke} on 2025/06/10 revealed that $\conc^\sharp_>=0.67$ and $\conc_<=0.56$ such that $\conc^\sharp_> > \conc_<$ and no quantum memory can be confirmed, even though readout-error mitigation was applied.

Nevertheless, more thought reveals that we do not need perfect reconstruction of the initial state after the application of $U^\dagger$, it suffices that the \emph{entanglement} regrows, which is a much weaker condition. Such a dynamics can be realized by coupling the first and second system qubit only to the first and second environmental qubit, respectively, via the unitary
\begin{align}
U_{\Sys_1 \Env_1}=U_{\Sys_2 \Env_2}=
    \begin{pmatrix}
        0&0&0&\iu\\
        1&0&0&0\\
        0&0&-\iu&0\\
        0&1&0&0
    \end{pmatrix}.
\end{align}
The unitary $U$ in Fig.~\ref{fig:2qubits_circuit} is thus chosen to be (apart from rearranging the order of the Hilbert spaces) given by $U=U_{\Sys_1 \Env_1} \otimes U_{\Sys_2 \Env_2}$. 
Note that the four-qubit unitary transformation used here is essentially a product of two two-qubit unitaries and the two system qubits never directly interact with each other. This choice is motivated by the observation that the more qubits in the actual architecture have to interact with each other, the more qubit pairs with average or bad connectivity have to be used. The latter leads to more noise and thus in turn prevents the quantumness of the memory to be witnessed. 

We will apply additional readout-error mitigation, which we did not do in the single-qubit case. Readout mitigation provided by IBMQ requires two additional circuits to be executed, such that we run 83 instead of 81 circuits. Those two ``empty'' circuits measuring all qubits in the $\ket{0}$-state and the $\ket{1}$-state, respectively, are used to gain information on the calibration of the quantum computer and help to reduce the readout-error. Note that this mitigation procedure is purely based on classical statistics. However, a quantum simulation on \texttt{ibm\_sherbrooke} on 2025/06/27 reveals that
\begin{align}
    \conc^\sharp_>(t_1) = 0.72 < 0.89 =\conc_<(t_2),
\end{align}
which shows that the property of this circuit of being only realizable with quantum memory is reflected properly by the quantum simulation on the noisy quantum computer. This can also be confirmed with respect to the entropic witness introduced in Ref.~\cite{BaeBeyStr2024} which is also suitable to characterize memory in higher-dimensional quantum dynamics.
To summarize, once the execution of the circuit is implemented within reasonable time and gate limits, and good interconnectivity between the involved qubits, contemporary quantum computers are capable of providing and witnessing quantum memory in non-Markovian quantum dynamics.

\section{Conclusions}
\label{sec:conclusions}

In this paper we studied non-Markovian quantum dynamics implemented on a contemporary quantum computer of IBM Quantum. We have shown that those quantum computers are capable of providing and also witnessing quantum memory in single- and two-qubit dynamics despite noise effects disturbing the dynamics. This is a crucial feature to simulate quantum dynamics on real-world devices and to exploit the advantage of quantum hardware over classical one.

We chose  to investigate a non-Markovian amplitude damping process which has previously been studied from a theoretical perspective and which turned out to require quantum memory \cite{BaeBeyStr2024}. This single-qubit dynamics can be modeled using another qubit as the environment and applying repeated unitary operations on the initially uncorrelated system-environment state.
For the implementation on the quantum computing resource \texttt{ibm\_sherbrooke} we furthermore added an ancilla qubit which was initialized to be in a maximally entangled state with the system.
Quantum state tomography on the system-ancilla state provides access to the Choi-Jamiołkowski state of the dynamics and together with the collision model approach allows for the investigation of the dynamics at discrete times.
The evolution of an entanglement measure with an ancilla can be regarded as a witness of non-Markovianity \cite{RivHuePle2010}. Similarly, comparing the concurrence of formation with the concurrence of assistance at an earlier time can be used to identify dynamics where the memory is necessarily quantum~\cite{BaeBeyStr2024}. 
Since we have access to the experimentally observed system-ancilla states we were able to compute the corresponding quantities which showed that the quantum computer is able to uphold quantum memory long enough to be witnessed.

Note that full quantum state tomography or full process tomography is not necessary for the quantum memory witness used here, according to Refs.~\cite{YuOhsNguNim2025, beyer2025onesidedwitnessquantumnessgravitational} it is possible to perform partial process tomography which is sufficient in certain cases. However, using the builtin quantum state tomography framework of IBMQ, it does not only return enough information to be used for a witness of quantum memory, but also the full quantum state which in addition provides information on other properties such as for example non-Markovianity criteria, which we also investigated here.
However, numerical investigations of the spin boson model have shown that the process tensor, which contains more information than a combination of two quantum maps, is more sensitive in detecting quantum memory \cite{BaeLinStr2025:p}. It may therefore be interesting to challenge contemporary quantum computers to implement full or reduced process tensor tomography in the two-qubit case and diagnose quantum memory from suitable multi-time quantities.

\section*{Acknowledgements}
We acknowledge the use of IBM Quantum services for this work. The views expressed are those of the authors, and do not reflect the official policy or position of IBM or the IBM Quantum team.
We thank Konstantin Beyer for valuable discussions and comments on the manuscript. Furthermore, C.~B.~acknowledges support by the German Academic Scholarship foundation.

\bibliography{literature}

\appendix

\begin{figure*}[htb]
    \centering
    \begin{quantikz}
         & \gate{H} & \ctrl{1} &  \\
         & \ghost{X}  & \targ{}&
    \end{quantikz}
    =\begin{quantikz}
         & \rzph & \rzph & \sx & \rzph & \rzph & \ecr &  & \\
          & & \rzph & \sx & \rzph & \rzmph & & \measure{+} & 
    \end{quantikz}
    \caption{Transpiled circuit for the initial Bell state of system and ancilla. While the global phase on the left-hand side is zero, the global phase on the right hand side is $\pi$ because this leads to a more efficient implementation when combined with the unitary from Fig.~\ref{fig:transpiled-u}.}
    \label{fig:transpiled-bell}
\end{figure*}
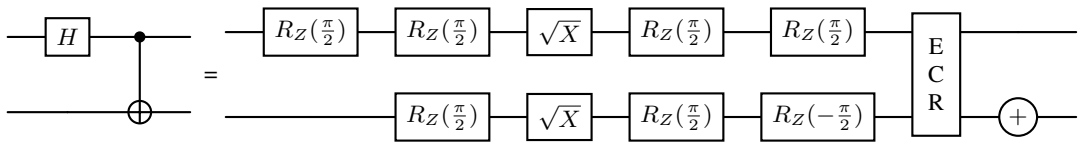

\begin{figure*}
    \centering
    \begin{quantikz}[column sep=0.2cm]
        & & \rzmpf & \sx & \ecr & \sx & \rztpf & \sx & \rzmp & \ecr & \rzmp & \sx & \rzmtpf\\
        & \rztpf & \sx & \rzmp & & \rzmp & \sx & \rztpf & \sx & & \sx & \rzpf
    \end{quantikz}
    \caption{Transpiled circuit for the unitary $\ug$ with $\gdt = \pi/4$ acting on system and environment.}
    \label{fig:transpiled-u}
\end{figure*}
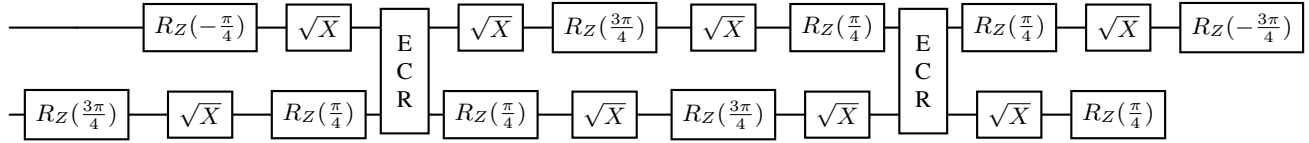

\section{Time-continuous limit of the collision model}
\label{sec:time_continuous_limit}
In order to obtain the local map for a single collision of time $\dt$ of system and environment, we need the global map described by the unitary $\ug$ from Eq.~\eqref{eq:ug} which reads
\begin{align}
    \map\left[\rho_{\Sys \Env}\right] = \ug \rho_{\Sys \Env} \ug^\dagger.
\end{align}
For very short times $\dt$ one can obtain a valid approximation $\mapser$ of the generator of the dynamics by expanding the map in a series up to first order in $\dt$ and calculating
\begin{align}
    \cpt = \text{tr}_{\Env}\left[\e^{\mapser \frac{\tau}{\dt}}\right],
\end{align}
with $\tau$ being the time.
Note that $\cpt$ does not include the evolution on the environment anymore, however the system dynamics depends on the initial state of the environmental qubit, we here choose it to be $\ket{0}$.
This exponentiation can be performed explicitly using the formalism of Bloch vectors $\vec{r}$ and the Bloch representation of quantum channels~\cite{BetSzaWer2002}. There the Bloch vector $\vec{r}$ is transformed linearly under the action of a map $\cpt$ such that
\begin{align}
    \vec{r}' = \cpt \vec{r},
\end{align}
and with $t=\g \tau$ being the dimensionless time the corresponding continuous-time map $\cpt$ becomes
\begin{align}
    \cpt_t = 
    \left(
    \begin{array}{cccc}
     1 & 0 & 0 & 0 \\
     0 & \cos (t) & 0 & 0 \\
     0 & 0 & \cos (t) & 0 \\
     \sin ^2(t) & 0 & 0 & \cos ^2(t) \\
    \end{array}
    \right).
\end{align}
Such a time-continuous qubit map can be transformed into a master equation by considering the generator
\begin{align}
    \mathcal{G} = \dot{\cpt}_t \circ \cpt_t^{-1}
\end{align}
and comparing coefficients of the matrix representation of this generator and a general Lindblad-type generator \cite{ZimSteBuz2005, ZimBuz2010, Kasatkin2023}. This leads exactly to the form of the master equation presented in Eq.~\eqref{eq:nMadthLindblad}.

\section{Implementation on IBM Quantum}
\label{sec:app-implementation}
The fundamental gates on \texttt{ibm\_sherbrooke} are the two single-qubit gates $\sqrt{\mathrm{X}}$ and $\mathrm{RZ}\left(\vartheta\right)$ and the two-qubit ECR gate. In order for a circuit to be executed on a quantum computer it has to be transpiled to be expressed in terms of the native gates only. We used the internal default algorithm provided by IBMQ yielding Figs.~\ref{fig:transpiled-bell} and~\ref{fig:transpiled-u}.

\end{document}